\documentstyle{article}
\def\ltsim{\raise 2pt \hbox {$<$} \kern-1.1em \lower 4pt \hbox {$\sim$}}
\def\gtsim{\raise 2pt \hbox {$>$} \kern-1.1em \lower 4pt \hbox {$\sim$}}
\begin{document}
\begin{center}
   {\LARGE  VLBI Observations of a Complete Sample of Radio \par
Galaxies V. 3C346 and 4C31.04: two Unusual CSS Sources. \par}
\vskip 10pt
{\large W. D. Cotton$^1$, L. Feretti$^{2,3}$, G. Giovannini$^{2,3}$, T.
Venturi$^2$,
L. Lara$^4$, \par
J. Marcaide$^5$, and A. E. Wehrle$^6$ \par}
\end{center}
\vskip 10pt
\par\noindent
$^1$National Radio Astronomy Observatory, 520 Edgemont Road,
Charlottesville, VA 22903-2475
\par\noindent
$^2$Istituto di Radioastronomia del CNR, Via P. Gobetti 101, I-40129
Bologna, Italy.
\par\noindent
$^3$Dipartimento di Astronomia, Universit\'a di Bologna, via Zamboni 33,
I-40100
Bologna, Italy.
\par\noindent
$^4$Instituto de Astrof\'{\i}sica de Andaluc\'{\i}a, CSIC, Apdo. 3004,
18080 Granada, Spain.
\par\noindent
$^5$Departamento de Astronom\'{\i}a, Universitat de Val\`encia,
46100 Burjassot, Spain.
\par\noindent
$^6$Infrared Processing and Analysis Center, California Institute of
Technology, M/S 100-22, Pasadena, CA 91125.

\vskip 10pt
Accepted for the publication in the {\it Astrophysical Journal}, October 20th
1995 issue

\vfill\eject

\centerline{\bf ABSTRACT}
\vskip 1em
\begin{quotation}

   We present observations at 1.7 and 8.4 GHz of two Compact Steep Spectrum
(CSS) sources from a complete sample of low-intermediate power radio galaxies.
3C346 shows an asymmetric structure with a one-sided ``jet'' and ``hot spot''.
Present observations suggest that the classification of this source as a CSS is
inappropriate, and that it is a common radio galaxy at a small angle
to the line of sight.
Its properties are in agreement with the predictions of unified
schemes models.
4C31.04 shows more complex structure with the possibility of a centrally
located flat spectrum core in between two close lobes. We suggest that this
source could be a low redshift Compact Symmetric Object.

\end{quotation}

\par\noindent
{\bf Subject Headings}: radio continuum: galaxies - Galaxies: jets -
Galaxies: nuclei
\vfill\eject
\vskip 5pt
{\centerline{ 1. INTRODUCTION}}

   In recent years, the parsec scale radio structure of high power radio
galaxies and quasars has been extensively studied using Very Long Baseline
Interferometry (VLBI). In order to test unified scheme models and understand
the nuclear properties of radio sources, a complete statistical study of well
defined samples, as well as detailed analyses of individual sources
are necessary.
In addition, to effectively determine the morphological characteristics of the
parsec scale structures, observations at multiple frequencies are needed to
distinguish the nature of the different components of these sources. In this
paper we present results on two sources, 3C346 and 4C31.04, taken from a larger
sample of radio galaxies (Giovannini {\it et al.} 1990). Nearly simultaneous
VLBI observations were obtained for them at 1.7 and 8.4 GHz.

VLBI mapping and analysis of all the radio galaxies in our sample is in
progress (see Giovannini {\it et al.} 1994 and references therein). The two
radio sources which we present in this paper are classified, on the basis of
their kiloparsec scale structures, as Compact Steep Spectrum (CSS) sources
(Peacock \& Wall 1982, Fanti {\it et al.} 1990). The origin and evolution of
this class of sources is not yet well understood. On the basis of statistical
arguments, Fanti {\it et al.}, (1990) argue that sources of this class are
predominantly intrinsically small rather than apparently small due to
projection effects. It is not yet clear whether these sources are the early
stages of more extended sources or whether the interstellar medium in the host
galaxies is so dense that the radio sources cannot escape from the confines of
the galaxy.

The detailed study of the two sources discussed here will add new
information about the objects in this class.
Additionally, it is a step towards  the completion of the VLBI study
of our sample of radio galaxies.

{\bf 3C346 (B 1641+174)} is identified with an elliptical galaxy with
redshift 0.162.
It shows, at arcsecond and sub-arcsecond resolution, a
compact core component from which a one sided jet emerges toward the east,
extending for $\sim$4.5".
An extended structure is present, both around the jet and on the
opposite side, for a total extent of $\sim$12".
The prominence of the compact core and the amorphous nature of the
extended structure suggest to Spencer {\it et al.} (1991) that this source is
a larger source viewed nearly end on.
Confirmation that this source does not belong in the CSS class will
improve the statistical analysis of that class by removing a non
member.
This result can also test the current unified schemes models since,
according to these models, a high power radio galaxy viewed at a small
angle with respect to the line of sight would appear as a quasar like
object.

   {\bf 4C31.04 (B 0116+319)} is identified with a nearby (z=0.059)
elliptical galaxy.
The small distance to this CSS source allows a detailed study of its
parsec scale
structure and a comparison of its properties with more distant and powerful
CSS sources. Wrobel \& Simon (1986) give 0.327 GHz VLBI
results finding the source resolved into a double separated by about 0.07".
This result makes 4C31.04 a good candidate for the new class of sources
named Compact Symmetric Objects (CSO, see Readhead {\it et al.} 1994 for a
recent review).  The astrophysical importance of these sources is
discussed in recent papers (see e.g. Wilkinson {\it et al.} 1994,
Readhead {\it et al.} 1995).
Since the number of known CSOs is very low and up to now all are very
powerful objects with a large redshift, identification of a nearby CSO
might open new possibilities for a more detailed study of the physical
properties of this class of sources.
H$_0$ = 100 km sec$^{-1}$Mpc$^{-1}$ and q$_0$ = 1 is assumed throughout the
paper.

\vskip 5pt
{\centerline{ 2. OBSERVATIONS AND DATA REDUCTION}}

    The sources under study were observed in 1992 January -- February
using 7 Very Long Baseline Array (VLBA) telescopes, one Very Large
Array (VLA) telescope and, when possible, the antennas of Medicina,
Noto and Madrid  (Deep Space Network, DSS65) in Europe (see Table 1).

Data were obtained with the MkII VLBI recording system and correlated on the
Caltech/JPL Block 2 correlator in Pasadena. Amplitude calibration was
carried out on the basis of measured system temperatures and assumed
antenna sensitivities and correlator efficiencies.
The gain calibration was tested by observing the two VLBI calibrator
sources 0133+476 and 1739+522 several times during the experiment.
Both sources were observed nearly simultaneously at the VLA in order
to derive their total flux density. 0133+476 was marginally resolved on
all the baselines used. Its measured flux density at 1.7 and 8.4 GHz was 1.38
Jy and 0.97 Jy, respectively. A self calibrated model of 0133+476 accounting
for the total flux density was used in the amplitude calibration. 1739+522 was
used only to calibrate the 3C346 8.4 GHz data set, and it turned out to be
slightly resolved only on the transatlantic baselines. Its total flux density
measured with the VLA at this frequency was 2.45 Jy.

\vskip 5pt
{\centerline {\it 2.1 3C346}}
3C346 was observed at 1.7 GHz using only the US telescopes, since due to
scheduling and technical problems the 3 European telescopes were not
available. The 8.4 GHz observations were made using all telescopes but DSS65.
At this frequency, the source was not detected on baselines to Medicina. The
uv coverages obtained on 3C346 are given in Figure 1a,b. Since the
antennas used at 8.4 GHz were mostly located in the southwestern United States
with two antennas in Europe, there is a large hole in the middle of the
coverage.

Editing of data, amplitude calibration and the initial passes of self
calibration were performed in the Caltech VLBI package (Pearson 1991). Fringe
fitting, final self calibration and analysis was performed in the NRAO
Astronomical Image Processing System (AIPS). Fringe fitting used the global
method of Schwab \& Cotton (1983) with a solution interval of 10 minutes.

The data at both frequencies were first model fitted by means of the program
MODELFIT in the Caltech Package. This stage proved to be essential, since it
allowed us to locate the more diffuse easternmost component and to estimate
its total flux density. The best models obtained with the model fitting at
each frequency were used to start the self calibration. The initial self
calibration iterations adjusted only antenna phases but the final iteration
adjusted antenna gains as well. We used a short solution interval (2 - 10
seconds) in the phase self calibration cycles and a longer solution interval
(2 hours) in the last self calibration iteration.

At 8.4 GHz only the ``core'' region could be seen in the full resolution image
(HPBW = 2.5 $\times$ 1 milliarcsec PA = 0$^{\circ}$). We therefore made a
lower resolution image (HPBW = 23.5 milliarcsec) by applying a taper in the uv
plane in order to reveal the lower surface brightness lobe regions. Parameters
relevant to the maps and source are given in Table 2. In this
table the total flux density (S$_T$) is obtained from
the sum of the CLEAN components, the
core flux density (S$_c$) is the peak brightness of the core (assuming
it to be much smaller than the beam) and the ``jet''
flux density is obtained from the integral over a region containing the
visible portion of the jet up to 2 arcseconds from the core. Since the methods
of measuring these components are different, the sum of the core and jet flux
densities do not equal the total but agree within the uncertainties.

\vskip 5pt
{\centerline {\it 2.2 4C31.04}}

The 1.7 GHz observations of 4C31.04 were made with the US telescopes and
Medicina. The uv coverage is given in Figure 2a. The source was
undetected on the baselines to Medicina.  As the calibrator source 0133+476,
located near the source and observed between the scans of 4C31.04, was
detected on all baselines, the non-detection was assumed to be due to
source resolution.

All post correlation analysis was performed in the NRAO AIPS package. Fringe
fitting used the global method of Schwab \& Cotton (1983) with a
solution interval of 12 minutes.
The self calibration iterative procedure for 4C31.04 was started using
a point source model.

The 8.4 GHz observations of 4C31.04 were made using all telescopes but
Medicina (see Table 1). The uv coverage obtained on 4C31.04 at 8.4
GHz is shown in Figure 2b. Fringe fitting used the model obtained
at 1.7 GHz. Since 4C31.04 is strongly resolved, the fringe fitting solutions
were smoothed with a median window filter and failed solutions were replaced
by interpolated values of good solutions. There were no detections of the
source on any baselines involving Noto or DSS65 and only occasional detections
involving the VLBA antennas at North Liberty and Brewster. Again, the
calibrator source was observed many times in between the source scans and
always detected on all baselines so the non detections are assumed to
be due to the resolution of 4C31.04.

Only phases were adjusted in the self calibration process, which was started
with the model derived from the 1.7 GHz observations. Since the source was
undetected on long baselines, we tapered the uv data using a Gaussian function
that drops to 30\% at 20 million wavelengths, and data from projected
baselines longer than 30 million wavelengths were excluded; the resulting
synthesized beam size is 14.7 $\times$ 6.6
milliarcseconds with a position angle of -16$^{\circ}$. This beam
is similar in size to that derived for 4C31.04 at 1.7 GHz.
Parameters relevant to the maps and the source are given in Table 3.
The last two columns in this table (``S$_E$'' and ``S$_W$'')
are the integrated flux densities of the eastern and western components as
measured by integrating over the relevant regions in the images.  The region
between the lobes was excluded; see Section 3.2 for a discussion about the
location of the core.

\vskip 5pt
{\centerline { 3. RESULTS}}
\vskip 5pt
{\centerline {\it 3.1 3C346}}
\vskip 5pt

   The images of 3C346 at 1.7 and 8.4 GHz are shown in Figure 3a,b,c. At
1.7 GHz we detect two components separated by $\sim$2.2", in agreement with
Rendong {\it et al.} (1991), the western component being the stronger and more
compact. It coincides with the core detected by
Spencer {\it et al.} (1991). The eastern component is a bright knot in the
asymmetric jet (Spencer {\it et al.} 1991), and is surrounded by resolved
extended emission.

At 8.4 GHz, only the nuclear emission is detected at full resolution. We
classify this emission as a core with a short one-sided jet,
oriented towards  the same side as the extended jet.
The eastern knot is completely resolved in
this map, but it is easily visible in the map produced at lower resolution
(HPBW = 23.5 mas). The jet-counter jet brightness ratio at 8.4 GHz is \gtsim
20 at $\sim$ 2 mas from the core.

Lower resolution images of Spencer {\it et al.} (1991), van Breugel {\it et
al.} (1992), and Akujor \& Garrington (1993) indicate that the structure shown
here is embedded in a larger halo of about 12" in
extent.  The compact ``core'' is located nearly at the center of this halo and
the ``knot'' is a bright spot in the asymmetric jet.

The high brightness of the core is clearly illustrated in Figure 3c
by its very small size at 8.4 GHz. Using the low resolution core flux density
at 8.4 GHz and assuming that the peak in the 1.7 and 8.4 GHz images
are coincident on the sky, then the apparent spectral index between these two
frequencies is $\alpha^{8.4}_{1.7}$ = -0.25. \footnote {S($\nu$) $\propto$
$\nu$ $^{-\alpha}$} 
The inverted spectrum is confirmed by data from the literature
(Rendong {\it et al.} 1991 and Spencer {\it et al.} 1991).
Our full resolution image at 8.4 GHz indicates that the emission in
the nuclear region is dominated by the inner portion of the jet;  the
peak brightness at the core corresponds to only 40\% of the core flux
density measured from the lower resolution image.
Since the core spectrum obtained from the low resolution images is
inverted, then the spectrum of the inner jet must be flat or inverted.

\par\noindent

\vskip 5pt
{\centerline {\it 3.2 4C31.04}}
\vskip 5pt
   The images of 4C31.04 at 1.7 and 8.4 GHz are shown in Figure 4a,b.
Due to the resolution of the source on the longer baselines, the final uv
coverage is similar at the two frequencies resulting in similar synthesized
beam sizes. The poor uv coverage at 8.4 GHz implies that the details in the
image may not be fully reliable. At 1.7 GHz the total flux density measured
with the VLA was 2.51 Jy and the final VLBI CLEAN image contained 2.46 Jy,
which is 98\% of the total. At 8.4 GHz the total CLEAN flux density in the
image was 0.79 Jy or 76\% of the measured total intensity of 1.04 Jy. The
missing flux density at 8.4 GHz is presumably in the larger size scales which
were better sampled at 1.7 GHz. The source appears as a double; the eastern
lobe is stronger and more compact than the western one which shows a distorted
structure.

   A faint bridge of emission is visible between the lobes at 1.7 GHz, while a
faint component is present at 8.4 GHz. Although the limited uv coverage and
sensitivity make the reliability of this 0.014 Jy component to be somewhat
uncertain, it is supported by the presence and morphology of radio emission in
this area at 1.7 GHz. In fact, this low frequency map clearly shows that the W
lobe, in addition to its extension in N-S direction, has an E-W extension,
i.e. in the same direction of the extension of the E lobe.
At 1.7 GHz, the flux density in this region is about 0.030 Jy, which would
imply a spectral index of $\approx$~0.5 or flatter, since the measurement at
the lower frequency probably includes some extended emission. Therefore this
feature could be the location of a flat spectrum core. We note that if such a
component were the core, then the ratio between the total radio power and the
core radio power for this source would be in agreement with the correlation
found for extended radio  galaxies by Giovannini {\it et al.} (1988).
This correlation ``predicts'' an unbeamed core flux density of about
17 mJy at 5 GHz, i.e. of the same order of magnitude of our core candidate.
This tends to support our identification of the core and suggests that
its flux density is only weakly affected by beaming effects, as is
expected from the symmetric morphology of this source.
However, it is not yet clear if the correlation found by Giovaninni
{\it et al.} (1988) for radio galaxies can be used for CSO sources
since the nature of
these sources is still uncertain; in particular, it is not known if they
are young, evolving sources.

Alternatively, the ``core'' of 4C31.04 could be embedded in one of the two
lobes of radio emission, or even in any region in our map if its flux density
were too low to be detected.  The nondetection of the source on long
baselines where the extended emission is completely resolved puts an
upper limit of approximately 100 mJy on the flux density of the core.
We find it unlikely that one of the two ``lobes'' is the core based on
component spectra as well as size.
If the brightest regions in the two images are assumed to be co-located
on the sky, their spectral indices are too steep ($\approx$~0.6 for the E
component and $\approx$~0.7 for the W one).
Moreover, both lobes are completely resolved by our longest baselines.
We cannot exclude the presence of a faint core component imbedded in one of
the two lobes (in this case the E one is favored, being stronger and
with a flatter spectrum).
In this case, the source structure would be highly asymmetric.
The spectra of the eastern and western components including data at 0.327
GHz from Wrobel \& Simon (1986) are shown in Figure 5.
The eastern component appears to become optically thick below about 1
GHz while the western component remains optically thin to 0.327 GHz.

No jet--like feature is visible in our maps and the two lobes are clearly
extended and cannot be identified as jets or bright knots in a jet structure.
Despite our incomplete uv-coverage, a jet--like structure should be easily
detected at a brightness of a tenth of mJy/beam.
This means that no Doppler boosted jet--like component is present in
this source.
We favor the hypothesis that this source shows a double structure,
with two extended (at VLBI resolution), somewhat distorted lobes and a
faint core emission in between.

   Wrobel \& Simon (1986) discuss the presence of a time variable component in
this source at 15 GHz, with variations of as much as 0.4 Jy. However, we note
that the radio variability is mostly due to one flux density
measurement in 1979 with a large uncertainty.
Marscher {\it et al.} (1979) classify this source as
a bursting radio source at 21 cm, but 4 different VLA observations in the time
range 1978 October - 1980 December show no significant flux density
variability. Due also to the observed low core flux density discussed
above, we think that the reported radio variability of this source is
hard to believe and, if confirmed, very peculiar.

\vskip 5pt
{\centerline { 4. DISCUSSION}}
\vskip 5pt
{\centerline {\it 4.1 3C346}}

{\bf 3C346 (B 1641+174)} is identified with a 17.2 magnitude galaxy with a
redshift of 0.16. The parent galaxy is in a double system with the compact,
north west nucleus coincident with the radio core (Dey \& van Breugel, 1994).
Spectroscopic data show that it has optical properties similar to those of
QSOs; it is included by Hewitt \& Burbidge (1991) in their ``optical catalog
of extragalactic emission--line objects similar to quasi--stellar objects'' and
it is classified as a weak emission line radio galaxy by Fabbiano {\it et al.}
(1984). X-ray emission was detected by the Einstein Observatory (Fabbiano {\it
et al.} 1984). Dey \& van Breugel (1994) observed excess ultraviolet light
from the region of the radio knot which they interpreted as synchrotron
emission from a hot spot at the end of the radio jet.

The total radio power at 408 MHz is 2.14 $\times$ 10$^{26}$ W/Hz, therefore
this source is in the same radio power range of extended FR-II radio galaxies
(see Fanaroff \& Riley 1974 for a definition of FR-I and FR-II radio
galaxies). The large scale radio structure has been discussed by Spencer {\it
et al.} (1991), Akujor {\it el al.} (1991), van Breugel {\it et al.} (1992),
and Akujor \& Garrington (1993).
It consists of a compact, core component from
which a jet emerges toward the east, extending with some wiggles and bright
knots for $\sim$4.5". An extended cocoon is present both around the jet and on
the opposite side, for a total extent of $\sim$12". This structure suggested to
Spencer {\it et al.} (1991) that this source is a large source viewed
nearly end on.
Rendong {\it et al.} (1991) made VLBI observations at
0.6 GHz and fitted a two component model with a separation of 2.2 arcsecond in
a position angle of $80^{\circ}$.

The presence of relativistic jets in strong radio sources as quasars
and FR-II radio galaxies is now widely accepted (see Antonucci, 1993
for a recent review).
This hypothesis is now also supported by recent results on FR-I radio
galaxies (Parma {\it et al.} 1994; Capetti {\it et al.} 1995) which
show that the jet asymmetry is large close to the core and reduces
gradually at larger distance as expected for intrinsically symmetric
jets in which velocity decreases with the distance from the core.
Since a milliarcsecond one-sided jet is
visible in our map in the same direction of the large scale one-sided jet
found by Spencer {\it et al.} (1991), we interpret the one-sided jet as due to
a symmetric structure affected by Doppler favoritism.
We will use the available data to constrain the possible values of the
intrinsic jet velocity and of the orientation of the radio source with
respect to the line of sight.

Following Giovannini {\it et al.} (1994), we can constrain the jet velocity
$\beta$ and orientation to the line of sight $\theta$ in three different ways.
The first method assumes that the milliarcsecond jets are intrinsically
symmetric, and boosted by Doppler favoritism.
{}From the jet to counter--jet brightness ratio R (\gtsim 20) we
obtain in this case a lower limit to $\beta cos\theta$ of 0.54 (see Giovannini
{\it et al.} 1994 and references therein for a more detailed discussion on
this method). The allowed region in the $\theta - \beta$ space is shown in
Figure 6.

  The second independent constraint may be derived using the prominence of
core radio power with respect to the total radio power. We will use the
correlation found by Giovannini {\it et al.} (1988). This correlation has
been derived with no selection by source size and can be used
for sources with quasar--like strong cores.
It allows the derivation of the beaming enhancement necessary to account for
the core prominence in the unified scheme models.
The measured core power of 3C346 is 22.4 times larger than
that expected from the core versus total radio power relation, leading to the
allowed region for $\theta - \beta$ again drawn in Figure 6 (see
Giovannini {\it et al.} 1994 for a more detailed discussion on this method).

The third method compares the X-ray detected emission with that expected
by the Self Compton Model (Ghisellini  {\it et al.} 1993, Marscher
1987). With available data, this method is inconclusive, since
it leads to a lower limit to the Doppler factor $\delta$ \gtsim 0.1.
This does not allow any constraint on the jet velocity or orientation
angle with respect to the line of sight.

{}From Figure 6 the allowed region for $\beta$ and $\theta$ is
$\theta$ $<$ 32$^{\circ}$ and $\beta$ $>$ 0.8, so we conclude that this source
is at a small angle to the line of sight. Unified scheme models predict that
sources with a steep spectrum (i.e. 'lobe dominated') and with an angle with
respect to the line of sight 10$^{\circ}$ $<$ $\theta$ $<$ 40$^{\circ}$ should
be classified as steep spectrum QSS or broad line radio galaxies (see e.g.
Ghisellini {\it et al.} 1993). This is in agreement with the optical
properties similar to those of QSOs reported in literature for 3C346 (see
Sect. 4.1), but it is in contrast with the lack of broad emission lines in its
spectrum.
However, Dey \& van Breugel (1994), infer a large optical extinction
($A_v > 8$) for this galaxy which could explain the lack of detection
of  the broad line region, and suggest that this region might only be
visible in the infrared.

The projected angular size of this source is $\sim$ 12" corresponding
to a linear size of 21.4 kpc.
An angle of $\theta$ $<$ 30$^{\circ}$ with respect to the line of
sight implies an intrinsic linear size of $>$ 43 kpc.
In a size-power diagram (see e.g. De Ruiter {\it et al.} 1990) a
source with the  radio power of 3C346 is expected to have a linear
size between 50 and 500 kpc.
Therefore, the deprojected lower limit on the size obtained for 3C346
($>$ 43 kpc) is  on the low end of the range but not peculiar.
This strongly suggests that 3C346 does not belong in the CSS class,
but is a common FR--II radio galaxy foreshortened by projection.
In this scenario, the ``knot'' visible at low resolution in our images
would correspond to the ``hot spot'' at the end of the jet.
Detection of superluminal motion in the core would further strengthen
this conclusion.

\vskip 5pt
{\centerline {\it 4.2 4C31.04}}

   {\bf 4C31.04 (B 0116+319)} is identified with the bright elliptical galaxy
MCG 5-4-18 (Caswell \& Wills 1967), a member of a close pair of galaxies, with
a redshift of z=0.059 (Heckman {\it et al.} 1983).
Van den Bergh (1970) reports [OII] emission in 4C31.04 and Heckman {\it et
al.} (1983) describe this source as showing low excitation forbidden oxygen
lines as well as a non stellar continuum optical and infrared source.
Mirabel (1990) reports the detection of HI absorption in this galaxy and a
high-velocity cloud of atomic hydrogen against the nuclear compact
continuum radio source at the galaxy center.
Marscher {\it et al.} (1979) failed to detect X-ray using HEAO 2.

The total radio power at 408 MHz is 1.26$\times$10$^{25}$ W/Hz i.e. in
the same range of the most powerful FR-I radio galaxies.
Perley (1982) indicates that the radio source is smaller than 1"
in extent.
Wrobel \& Simon (1986) presented a 0.327 GHz VLBI map where the source
is a double separated by about 0.07".
Kulkarni \& Romney (1990) give results from a three
antenna VLBI array at 1.4 GHz also giving a double with a separation
of 0.07".

    The parsec scale structure of this source consists of two almost equal
flux density components, which are resolved by the longest baselines. These
show some distortion, but the overall structure is rather symmetric.    This
source is clearly different from most of the VLBI sources found in the
literature and in our sample (see Giovannini {\it et al.} 1994, and references
therein), in which a core-jet structure is dominant. Both components have a
clearly extended radio structure which do not satisfy the generaly accepted jet
definition (Bridle \& Perley, 1984). They resemble two extended lobes very
similar to those found in radio galaxies at a much larger scale.  The
interpretation of the morphology of 4C31.04 is not straightforward.  Sanghera
\& Spencer (1993) suggest that strong interactions with the ISM may explain
the small size and distorted appearance of many CSS sources associated
with quasars.

The line emission reported by Vanden Bergh (1970) and  Heckman {\it et
al.} (1983) are relatively weak compared to other radio galaxies with
compact radio sources.
Therefore, there is no compelling evidence of a strong interaction of
the radio jets with the ambient medium.
The apparent small size could then be due simply to the source being young,
but this alone does not explain its rather distorted appearance.
Another possibility is to invoke some other effects such as
precession, although the source does not show the symmetry expected from such
a model, or from some simple motion of the core.

   Another possible cause of the apparent distortion of the structure of
4C31.04 could be gas flow inside the galaxy as is reported in NGC4874 by
Feretti \& Giovannini (1985) on a larger scale. Gas infalling from tidal
interactions with the other galaxy of this double system could provide
systematic gas flow in the nuclear region.
Mirabel (1990) reports observations of HI inflow in absorption against
the continuum source which indicate a considerable amount
of HI gas in the central region of this giant elliptical galaxy.
Normal elliptical galaxies are rather poor in interstellar gas so
infalling gas could result in systematic gas motions even at the nucleus.

The symmetric  radio structure of this source is reminiscent of that of the
CSOs (see Readhead {\it et al.} 1994 for a recent review), which are very
luminous, small, and short-lived objects. Prototypes of this class are
0108+388 (Conway {\it et al.} 1994), 0710+439 and 2352+495 (Conway {\it et
al.} 1992, Wilkinson {\it et al.} 1994). In table 4 we compare the
physical parameters of 4C31.04 and 2352+495.
4C31.04 is at a much lower redshift and has a lower radio
power, linear size and equipartition magnetic field strength than
2352+495.
The equipartition minimum pressure within the radio emitting regions,
0.63 U$_{min}$ (see Feretti {\it et al.} 1992), is lower
than that of canonical CSOs, therefore the two lobes could be
statically confined by the external pressure  present in the Narrow
Line region.

   In a discussion of CSOs, Wilkinson {\it et al.} (1994) point out that the
well studied CSOs 2352+495 and 0710+439 have very strong similarities: both
are identified with galaxies with narrow emission lines and not broad emission
lines. Host galaxies have distorted isophotes and nearby companions. The radio
emission has a low percentage of polarized flux density.
The host galaxy of 4C31.04 is a member of a double galaxy system and
its optical properties are similar to the CSOs host galaxies.
Moreover, the 8.4 GHz polarization measured using the
VLA during calibration observations give a low polarization of
$0.56\pm0.02\%$.
These properties of 4C31.04 strengthen the argument for its
interpretation as a symmetric source.
We conclude that 4C31.04 could represent a nearby and faint member of
the CSO class.

\vskip 5pt
{\centerline { 5. CONCLUSIONS}}

   The prominence of the high brightness, compact core seen in 3C346
together with the large scale structure of the radio emission and
absence of a counter jet are evidence that this source is viewed nearly end
on and may not be confined to its host galaxy. The optical properties of the
host galaxy are in agreement with the expectations of unified scheme
models.

4C31.04 is a more difficult case to analyze. There is not a prominent flat
spectrum high brightness ``core'' at the epoch of these observations and no
jet--like feature  is evident in our maps.  This would argue against Doppler
boosting due to a relativistic jet being significant and thus against the
source being seen end on. Although with the current data it is not possible to
firmly establish the core location, we favor the hypothesis that the core is
located between the main components and that this source has relatively
symmetric structure. The morphology and physical properties suggest that this
source could be a low redshift Compact Symmetric Object. The existence of a
nearby member of this class of objects is very interesting and, even if more
data are necessary to confirm the absence of relativistic motions and to
confirm the core identification in 4C31.04, its physical properties suggest
that CSO sources might be present also at our epoch and not only in high power
and high redshift sources.
Therefore, it has to be taken into account in future statistical tests
studing the cosmological evolution of CSOs and the possible
connection between CSOs and extended radio galaxies.

   If 4C31.04 is not affected by projection, then the distortions of the
structure must be intrinsic.
These could be due to ram pressure of a high velocity gas flowing into
the nuclear region of the parent galaxy as observed by Mirabel (1990).

\vskip 5pt
{\centerline { Acknowledgements}}
We would like to thank A.C.S. Readhead for useful discussions and the
staffs of the observatories who participated in
these observations and especially to the staff of the Cal Tech VLBI correlator.
The National Radio Astronomy Observatory is operated by Associated
Universities, Inc. under a cooperative agreement with the National Science
Foundation.

\vfill\eject
{\centerline {\large References}}
\par\noindent
Akujor, C. E., \& Garrington, S. T. 1993 in Sub-arcsecond Radio
Astronomy,
\par\parindent 30pt
ed. R. J. Davis \& R. S. Booth (Cambridge: Cambridge University
Press), p. 275

\par\noindent
Akujor, C. E., Spencer, R. E., Zhang, F. J., Davis, R. J., Browne, I.
W. A., and Fanti, C.
\par\parindent 30pt
1991, {\it M.N.R.A.S} {\bf 250}, 214

\par\noindent
Antonucci, R. R. J. 1993, {\it Ann. Rev. Astr. Ap.} {\bf 31}, 473

\par\noindent
Capetti, A., Fanti, R., Parma, P. 1995 {\it Astr. Ap.} in press

\par\noindent
Caswell, J. L., \& Wills, D. 1967, {\it M.N.R.A.S.} {\bf 135}, 231

\par\noindent
Conway, J.E., Myers, S.T., Pearson, T.J., Readhead, A.C.S., Unwin S.C.,
 and Xu, W.
\par\parindent 30pt
1994, {\it Ap. J}, in press

\par\noindent
Conway, J.E., Pearson, T.J., Readhead, A.C.S., Unwin, S.C., Xu, W., and
Mutel, R.L.
\par\parindent 30pt
1992, {\it Ap. J.} {\bf 396}, 62

\par\noindent
De Ruiter, H.R., Parma, P., Fanti, C., and Fanti, R. 1990, {\it Astr. Ap.}
{\bf 227}, 351

\par\noindent
Dey, A., \& van Breugel, W. J. M., 1994, {\it Astron. J.}, {\bf 107}, 1977

\par\noindent
Fabbiano, G., Miller, L., Trinchieri, G., Longair, M., and Elvis, M.
1984, {\it Ap. J.}, {\bf 277}, 115

\par\noindent
Fanaroff, B. L., \& Riley, J. M. 1974 {\it M.N.R.A.S.} {\bf 167}, 31

\par\noindent
Fanti, R., Fanti, C., Schilizzi, R. T., Spencer, R. T., Rendong, N.,
Parma, P.,
\par\parindent 30pt
van Breugel, W. J. M., and Venturi, T. 1990, {\it Astr. Ap.} {\bf
231}, 333

\par\noindent
Feretti, L., \& Giovannini, G. 1985, {\it Astr. Ap.} {\bf 147}, L13

\par\noindent
Feretti, L., Perola, G. C., and Fanti, R. 1992, {\it Astr. Ap.} {\bf 265}, 9

\par\noindent
Giovannini, G., Feretti, L., Gregorini, L., and Parma, P. 1988,
{\it Astr. Ap.} {\bf 199}, 73

\par\noindent
Giovannini, G., Feretti, L., and Comoretto, G. 1990, {\it Ap.J.}, {\bf
358}, 159

\par\noindent
Giovannini, G., Feretti L., Venturi, T., Lara, L.,
Marcaide, J., Rioja, M., Spangler, S.R.,
\par\parindent 30pt
and Wehrle, A.E., 1994, {\it Ap. J.}, in press

\par\noindent
Ghisellini, G., Padovani, P., Celotti, A., and Maraschi, L.,
1993, {\it Ap. J.} {\bf 407}, 65

\par\noindent
Heckman, T. M., Lebofsky, M. J., Rieke, G. H., and van Breugel, W., 1983,
\par\parindent 30pt
{\it Ap. J.}, {\bf 272}, 400

\par\noindent
Hewitt, A., \& Burbidge, G.,  1991, {\it Ap.J.Suppl.}, {\bf 75}, 297

\par\noindent
Kulkarni, V. K., \& Romney, J. D. 1990 in Compact Steep-Spectrum and
GHz-Peaked
\par\parindent 30pt
Spectrum Radio Sources, eds. C. Fanti, R. Fanti, C. P. O'Dea and R. T.
\par\parindent 30pt
Schilizzi, CNR, Bologna, 85

\par\noindent
Marscher, A. P., Marshall, F. E., Mushotzky, R. F., Dent, W. A.,
Balonek,
\par\parindent 30pt
T. J., and Hartman, M. F. 1979, {\it Ap. J.}, {\bf 233}, 498

\par\noindent
Marscher, A.P., 1987, in {\it Superluminal Radio Sources}, J.A. Zensus \&
T.J. Pearson Eds.,
\par\parindent 30pt
Cambridge University Press, p. 280

\par\noindent
Mirabel, I.F., 1990, {\it Ap.J.(Letters)}, {\bf 352}, L37

\par\noindent
Parma, P., de Ruiter, H. R., Fanti, R., Laing, R. 1994 in {\it The First
Stromlo
Symposium:
\par\parindent 30pt
The Physics of Active Galaxies}, G.V. Bicknell, M.A. Dopita,
\par\parindent 30pt
P.J. Quinn Eds., p. 241

\par\noindent
Peacock, J. A., \& Wall, J. V. 1982, {\it M.N.R.A.S.} {\bf 198}, 843

\par\noindent
Pearson, T.J., 1991, {\it BAAS}, {\bf  23}, 991

\par\noindent
Perley, R. A. 1982, {\it A. J.}, {\bf 87}, 859

\par\noindent
Readhead, A.C.S., Xu, W., Pearson, T.J., Wilkinson, P.N., and Polatidis, A.G.
\par\parindent 30pt
1994, in Compact Extragalactic Radio Sources, Ed. J.A. Zensus \& K.I.
\par\parindent 30pt
Kellerman, p. 17

\par\noindent
Readhead, A.C.S., Xu, W., Pearson, T.J., Wilkinson, P.N., and Polatidis, A.G.
\par\parindent 30pt
1995, in preparation

\par\noindent
Rendong, N., Schilizzi, R. T., Fanti, C., and Fanti, R. 1991,
{\it Astr. Ap.} {\bf 252}, 513

\par\noindent
Sanghera, H. S. \& Spencer, R. E. 1993 in Sub-arcsecond Radio
Astronomy,
\par\parindent 30pt
ed. R. J. Davis \& R. S. Booth (Cambridge: Cambridge University
Press), 367

\par\noindent
Schwab, F. R., \& Cotton, W. D. 1983 {\it A.J.}, {\bf 88}, 688

\par\noindent
Spencer, R. E., Schilizzi, R. T., Fanti, C., Fanti, R., Parma, P., van
Breugel,
\par\parindent 30pt
W. J. M., Venturi, T., Muxlow, T. W. B., and Rendong, N. 1991, {\it
M.N.R.A.S} {\bf 250},
\par\parindent 30pt
225

\par\noindent
Van Breugel, W. J. M.,  Fanti, C., Fanti, R., Stanghellini, C., Schilizzi,
R. T., and Spencer,
\par\parindent 30pt
R. E. 1992, {\it Astr. Ap.} {\bf 256}, 56

\par\noindent
Van den Bergh, S. 1970 {\it Pub. A. S. P.} {\bf 82}, 1374

\par\noindent
Wilkinson, P. N., Polatidis, A. G., Readhead, A. C. S., Xu, W., and Pearson,
T. J.
\par\parindent 30pt
1994, {\it Ap. J.(Letters)}, {\bf 432}, L87

\par\noindent
Wrobel, J. M., \& Simon, R. S. 1986, {\it Ap.J.}, {\bf 309}, 593

\par\noindent
\par\parindent 30pt

\vfill\eject

\begin{figure}
\caption{
a) The uv coverage for 3C346 at 1.7 GHz.
\hfill\break
b) The uv coverage for 3C346 at 8.4 GHz.
In both figures only the final data used to obtain the published maps
are drawn, points deleted as bad by editing or in the self calibration
cycles are not shown.
}
\label{3c346uv}
\end{figure}

\begin{figure}
\caption{
a) The uv coverage for 4C31.04 at 1.7 GHz.
\hfill\break
b) The uv coverage for 4C31.04 at 8.4 GHz.
In both figures only the final data used to obtain the published maps
are drawn, points deleted as bad by editing or in the self calibration
cycles are not shown.
}
\label{4c3104uv}
\end{figure}

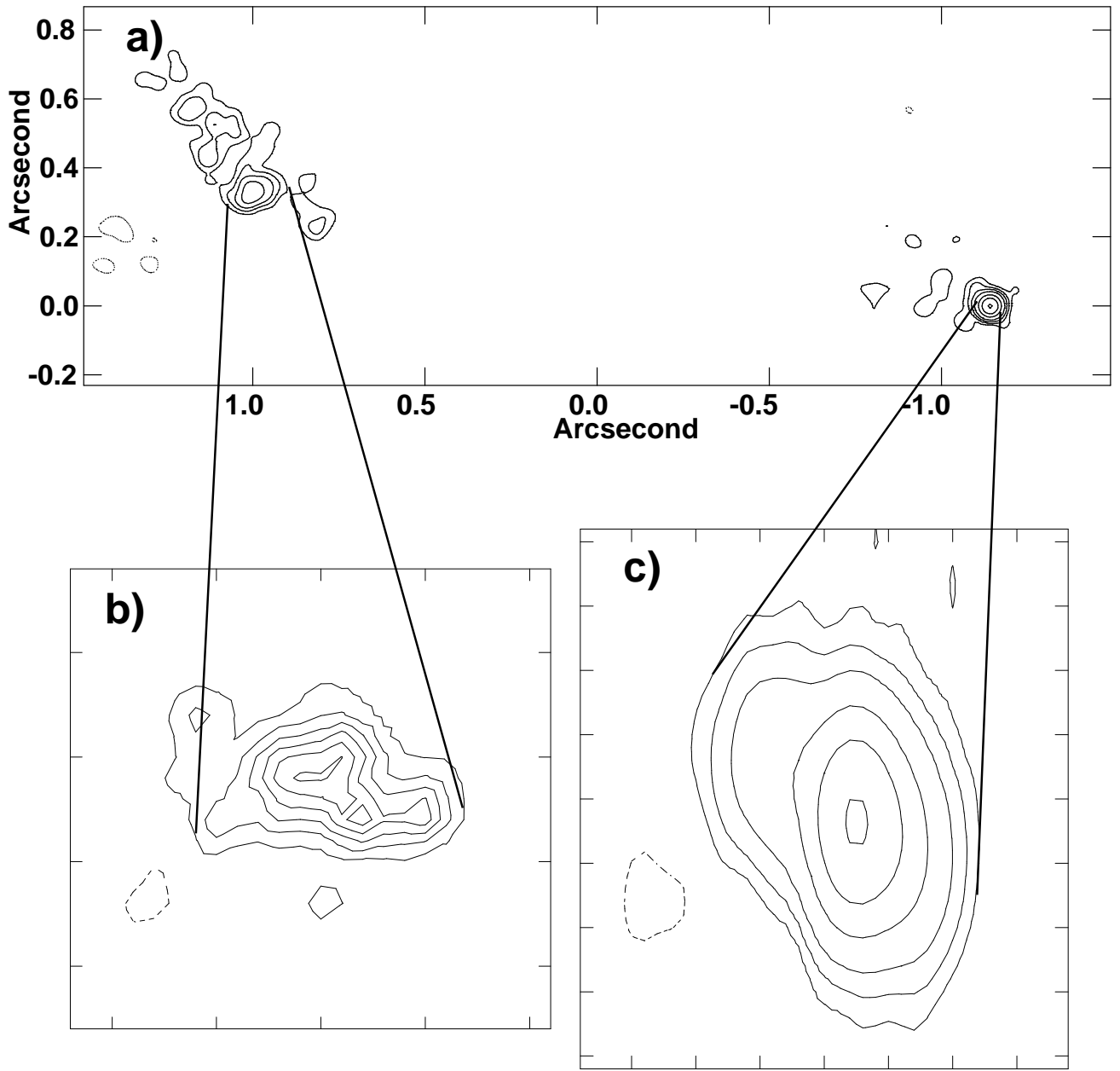
\begin{figure}
\caption{
a) 3C346 at 1.7 GHz.  The peak flux density is 149 mJy and the
contours are at -1, 1, 2.5, 5, 10, 25, 50, and 95 percent of the
peak.  The RMS in a empty region of the image is 0.25 mJy.
The restoring beam is 45 $\times$ 45 milliarcseconds.
\hfill\break
b) Low resolution image of ``knot'' in 3C346 at 8.4 GHz; tick marks
are every 50 milliarcseconds.
The peak  flux density is 6.2 mJy/beam and the contours are at -1,
1, 2, 3, 4, 5, and 6 mJy.
The RMS noise is 0.25 mJy and the restoring beam is 23.5 $\times$ 23.5
milliarcseconds.
\hfill\break
c) Full resolution image of ``core'' of 3C346 at 8.4 GHz; tick marks
are every milliarcsecond.
The peak  flux density is 86 mJy and the contours are at -2.5, 2.5,
5, 10, 25, 50, and 95 percent of the peak.
The RMS noise is 1.2 mJy and the restoring beam is 2.5 $\times$ 1
milliarcsecond at a position angle of 0$^o$.
}
\label{3C346}
\end{figure}

\begin{figure}
\caption{
a) 4C31.04 at 1.7 GHz.  The peak flux density is 472 mJy and the
contours are at -1, 1, 3, 5, 7, 10, 20, 30, 50, 70, 90 percent of the
peak.  The RMS in a empty region of the image is 1.5 mJy.
The restoring beam is 17 $\times$ 10 milliarcseconds at a position
angle of 72$^o$.
The cross marks the approximate location of the possible ``core'' seen
in b).
\hfill\break
b) 4C31.04 at 8.4 GHz on the same scale as a).  The peak flux density
is 201 mJy and the contours are at -3, 3, 5, 7, 10, 20, 30, 50, 70, 90
percent of the peak.
The RMS noise is 1.9 mJy.
The restoring beam is 15 $\times$ 7 milliarcseconds at a position angle
of -16$^o$.
}
\label{4C3104}
\end{figure}
\begin{figure}
\caption{
  The integrated spectra of the eastern and western components of
4C31.04.
The 0.327 GHz measurements are from Wrobel and Simon (1986); the other
values are derived from integrals over the relevant components in the
images presented here.  The region between the two major components is
not included.
The lines are shown to distinguish the two components and are not fitted
spectra.
}
\label{4c3104spec}
\end{figure}

\begin{figure}
\caption{
This figure shows the jet velocity, beta ($\beta$), versus orientation to the
line of sight, theta ($\theta$), plot for 3C346.
Curved lines marked ``A'' are derived from the observed core dominance
and the line marked ``B'' from the jet / counter-jet ratio.  The
disallowed regions are marked with diagonal stripes.
}
\label{allowed}
\end{figure}

\vfill\eject
\begin{table} [t]
\begin{center}
\caption{Observations}
\medskip
\begin{tabular}{lclll}
\hline \hline

Source&  Frequency & Stations &  Obs. Date & Obs. Time \\
\\
      &   GHz  & & & (hours) \\
\\
\hline
3C346   & 1.7 & Y1,PT,KP,LA,FD,NL,BR,OV      &  1992 Jan 31         &
6.5 \\
\\
3C346   & 8.4 & Y1,PT,KP,LA,FD,NL,BR,OV,L,N  &  1992 Jan 24         &
6.5 \\
\\
4C31.04 & 1.7 & Y1,PT,KP,LA,FD,NL,BR,OV,L    &  1992 Jan 30, Feb 2  &
9.0 \\
4C31.04 & 8.4 & Y1,PT,KP,LA,FD,NL,BR,OV,M,N  &  1992 Jan 23-25      &
9.0 \\
\\
\hline
\label{obstab}
\end{tabular}
\medskip \\

\end{center}
{\em Notes:}
L: 32m, Medicina (Italy); M: 34m, DSS65, Robledo (Spain); N: 32m, Noto (Italy);
Y1: 25m one telescope VLA (USA); PT: 25m, VLBA-Pie Town (USA);
KP: 25m, VLBA-Kitt Peak (USA); LA: 25m, VLBA-Los Alamos (USA);
FD: 25m, VLBA-Fort Davis; NL: 25m, VLBA-North Liberty;
BR: 25m, VLBA-Brewster; OV: 25m, VLBA-Owens Valley


\begin{center}
\caption{Map and Source parameters for 3C346}
\medskip
\begin{tabular}{lcccccccc}
\hline \hline
Source & Freq. & HPBW & (PA) & noise &  S$_T$ & S$_c$ &  S$_{jet}$ &
PA$_{jet}$ \\
\\
       & GHz   & mas  & deg & mJy/beam & mJy & mJy  &   mJy  &  o \\
\\
\hline
3C346  & 1.7 & 45x45  &  -  & 0.25   & 361 & 149 & 215 & 81 \\
\\
3C346  & 8.4 & 23.5x23.5& -  & 0.25   & 252 & 223 & 42 & 81 \\
\\
3C346  & 8.4 & 2.5x1   & 0  & 1.2    & 143 &  86 & & \\
\hline
\label{3c346tab}
\end{tabular}
\medskip \\
\end{center}
\end{table}

\vfill\eject

\begin{table} [t]
\begin{center}
\caption{Map and Source parameters for 4C31.04}
\medskip
\begin{tabular}{lcccccccc}
\hline \hline
Source & Freq. & HPBW & (PA)& noise  & S$_T$&  S$_c$ &    S$_E$  &   S$_W$ \\
\\
       & GHz   & mas  & deg &mJy/beam& mJy & mJy   &  mJy  & mJy \\
\\
\hline
\\
4C31.04 & 1.7 & 17x10 &   72 & 1.5  &  2460 &  ($<30$) & 1366 & 961 \\
\\
4C31.04 & 8.4 & 15x7  &  -16 & 1.9  &   790 &  (14) & 541 & 264 \\
\hline
\label{4C3104tab}
\end{tabular}
\medskip \\
\end{center}


\begin{center}
\caption{Physical Parameters of Selected CSOs}
\medskip
\begin{tabular}{llccccc}
\hline \hline
Name & & z & P$_T^{408} $ & Size & U$_{min}$  & H$_{eq}$ \\
\\
     & &   & W/Hz       & pc  & $10^{-6}$ erg/cm$^3$ & $10^{-3}$ gauss \\
\\
\hline
\\
4C31.04 & & 0.057 & $1.3\times 10^{25}$ & 70 & \\
\\
 & East & & & & 0.4 & 2.1 \\
\\
 & West & & & & 0.3 & 1.9 \\
\\

2352+495 & & 0.237 & $1.6\times 10^{26}$ & 150 & \\
\\
 & North Lobe & & & & 3 & 6 \\
\\
 & South Lobe & & & & 2 & 5 \\
\hline
\label{parmtab}
\end{tabular}
\medskip \\
\end{center}
{\em Note:}
2352+495 data are from Readhead {\it et al.} 1995.
\end{table}
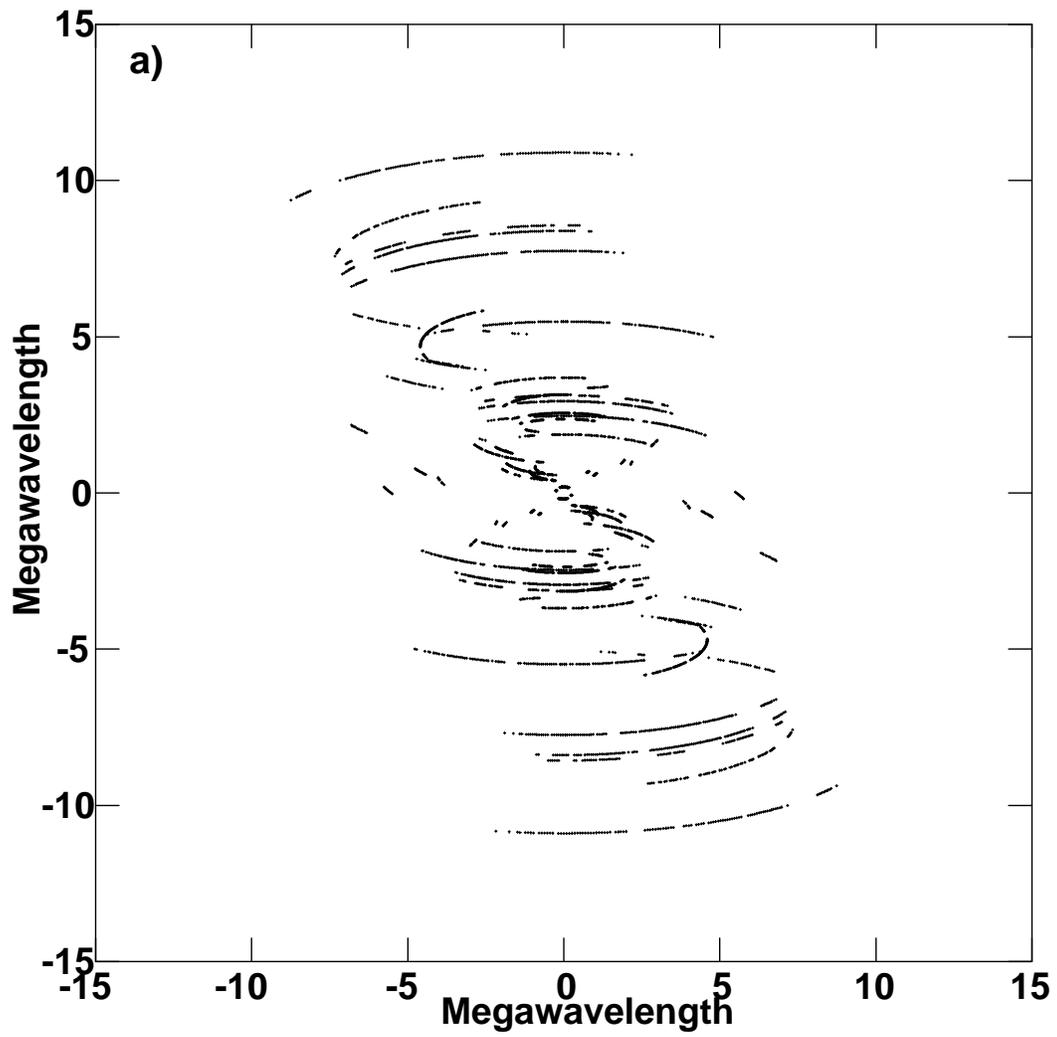
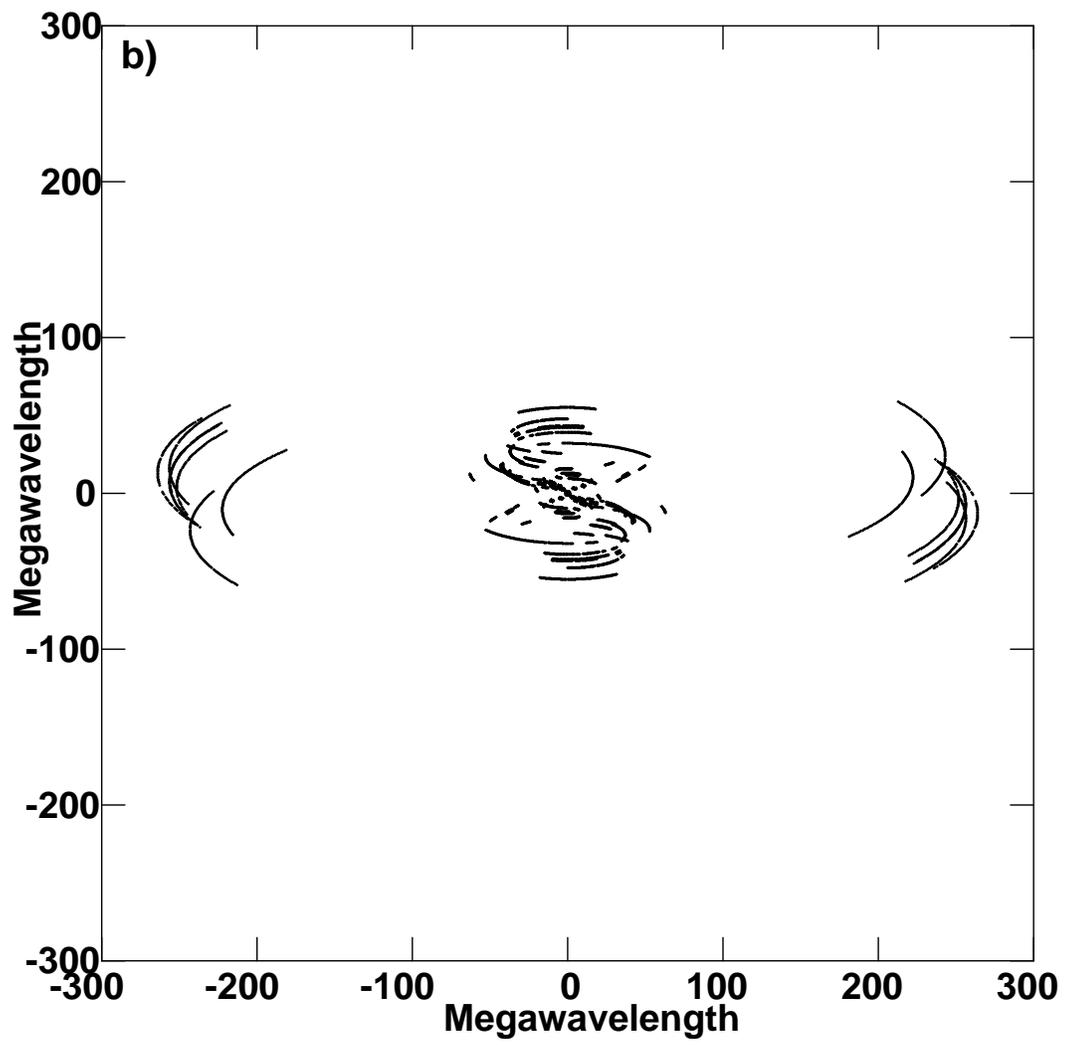
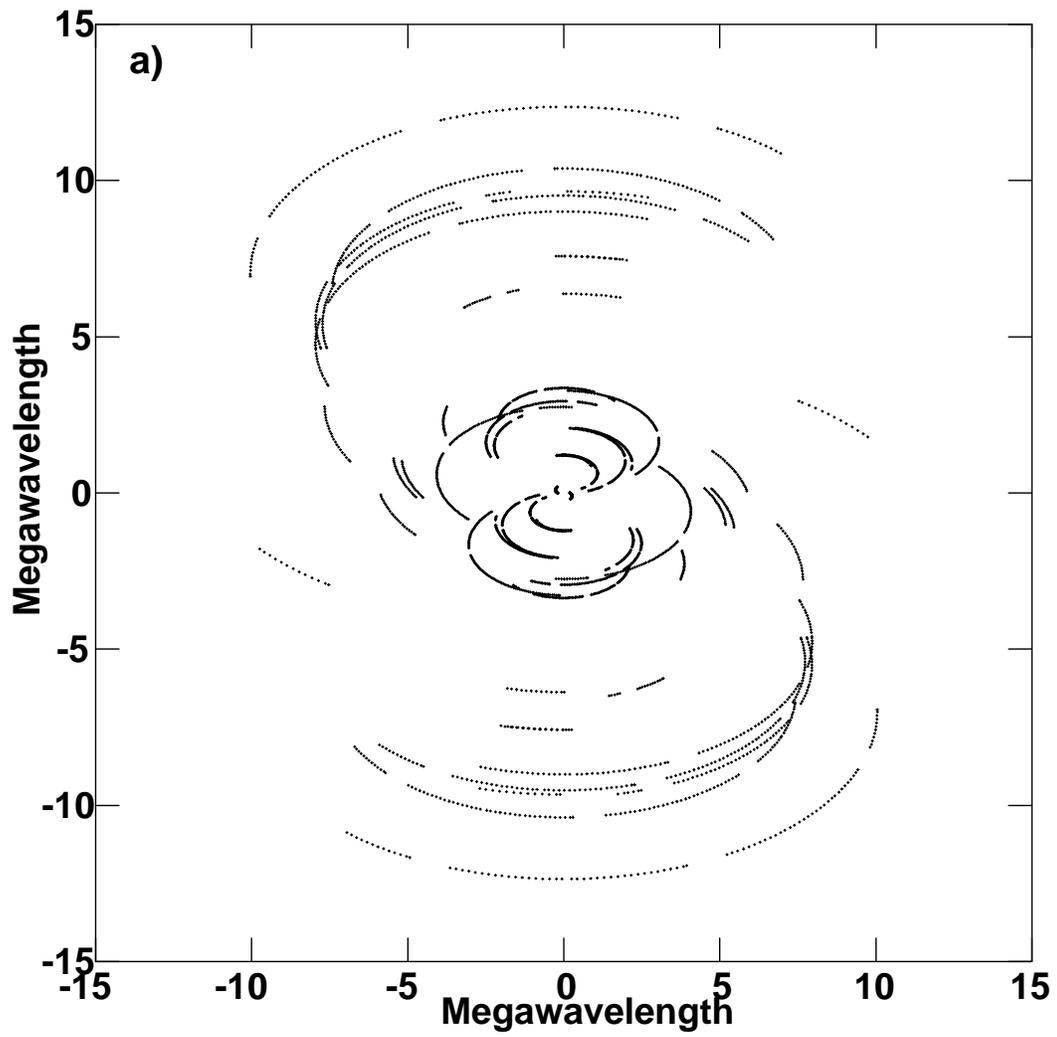
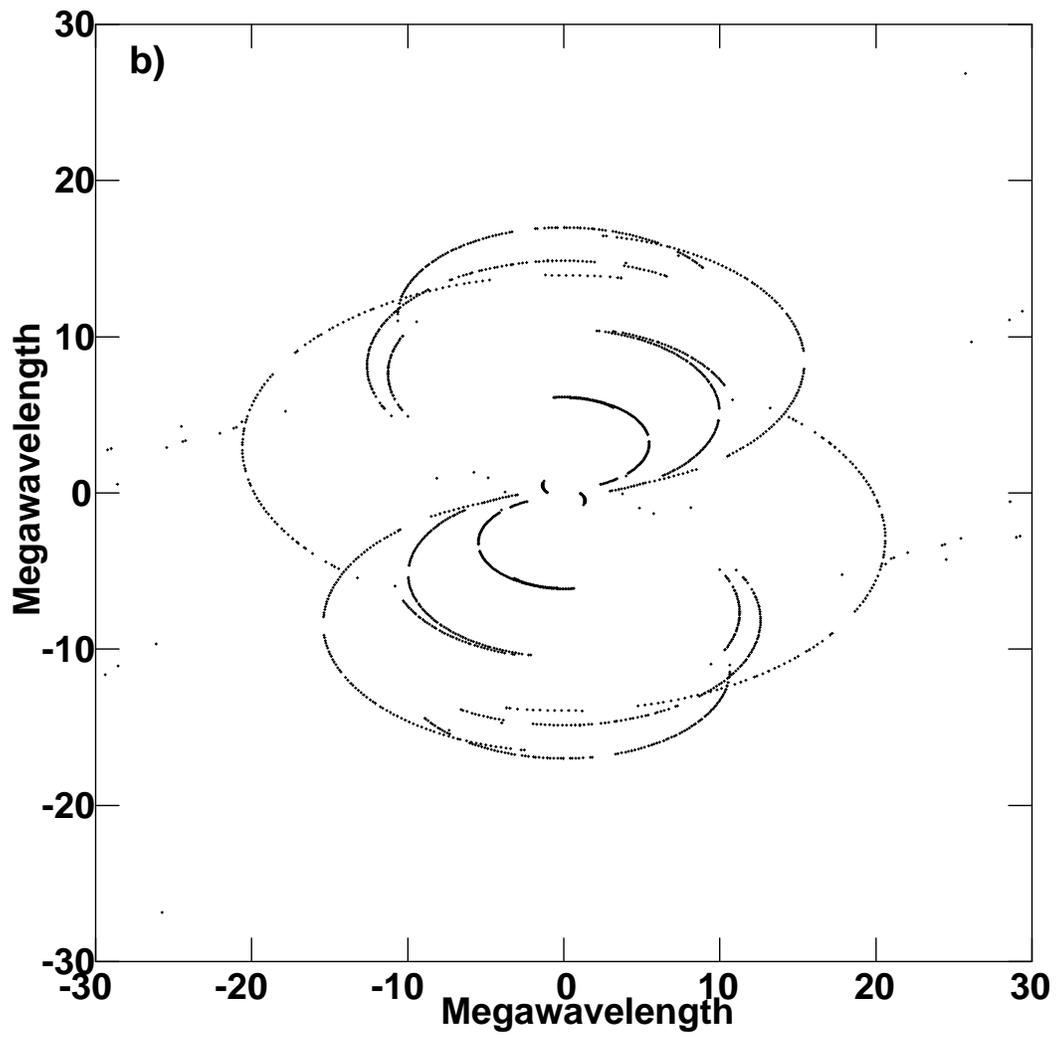
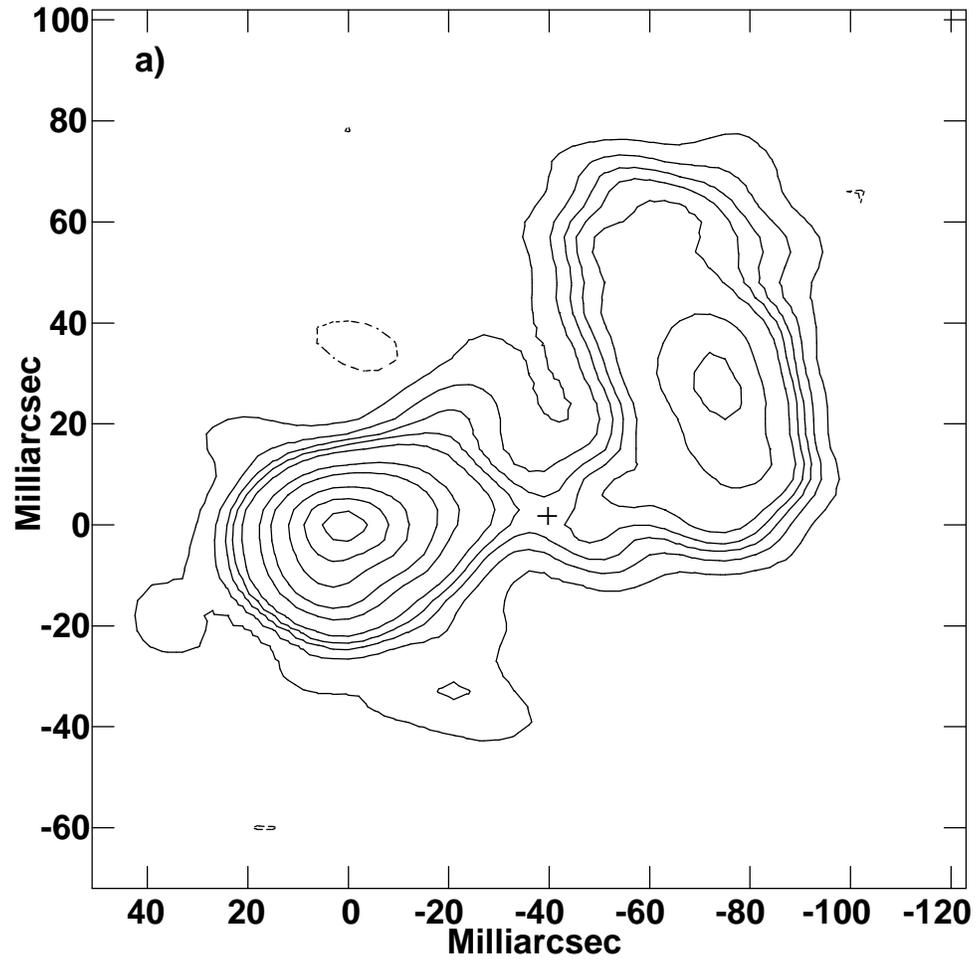
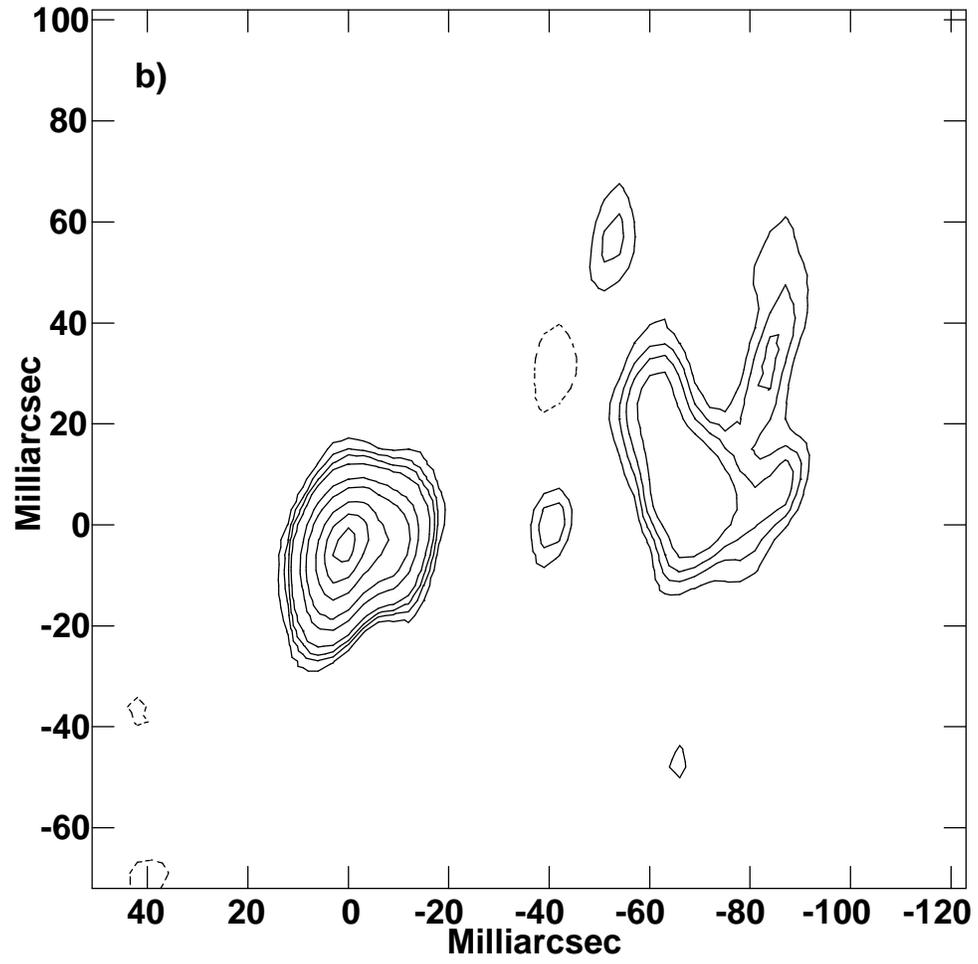
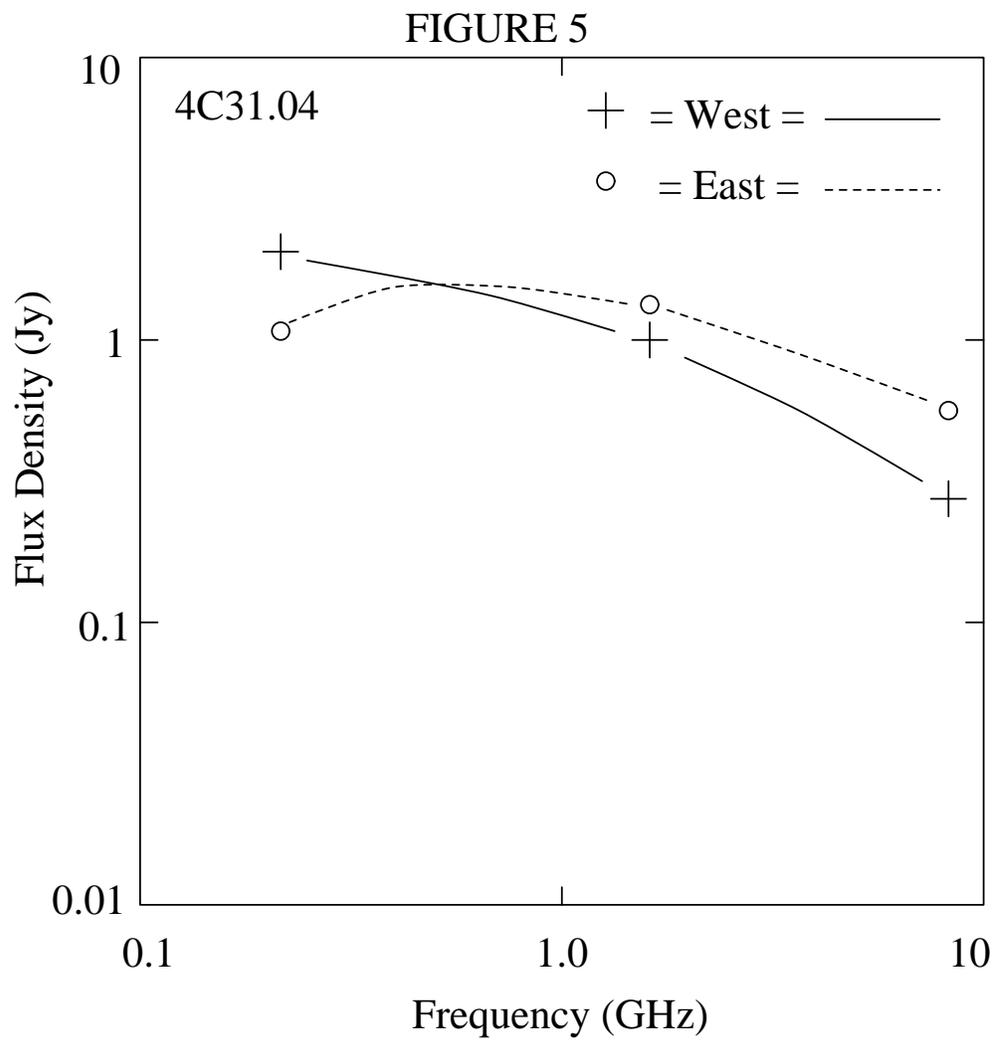
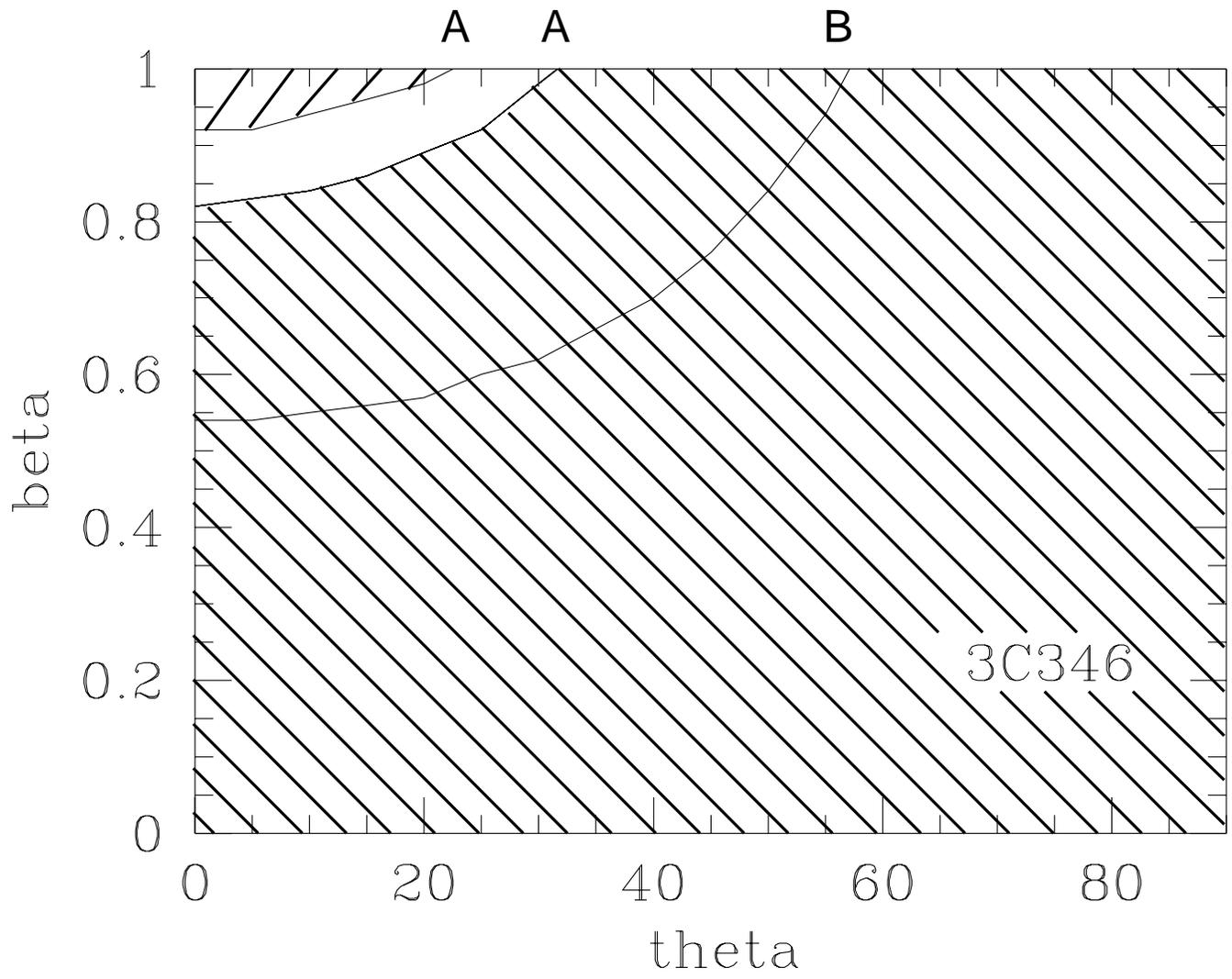
\end{document}